\documentclass{sigcomm-alternate}
\usepackage{amsfonts,latexsym,amsmath,amstext,amssymb,verbatim,epsfig,psfrag}
\usepackage{colordvi,pstcol}
\usepackage{graphicx,psboxit}
\usepackage{psfrag}

\begin{document}
\title{D-iteration based asynchronous distributed computation}

\numberofauthors{1}
\author{
   \alignauthor Dohy Hong\vspace{2mm}\\
   \affaddr{Alcatel-Lucent Bell Labs}\\
   \affaddr{Route de Villejust}\\
   \affaddr{91620 Nozay, France}\\
   \email{\em dohy.hong@alcatel-lucent.com}
}

\date{\today}
\maketitle

\begin{abstract}
The aim of this paper is to explain how the D-iteration can be used for an efficient asynchronous distributed computation. 
We present the main ideas of the method and illustrate them through very simple examples.
\end{abstract}
\begin{psfrags}
\section{Introduction}\label{sec:intro}
As an improved or alternative solution to existing iterative methods
(cf. \cite{Golub1996, Saad, Bagnara95aunified}),
the D-iteration algorithm has been proposed in \cite{d-algo} in a general context
of linear equations to solve $X$ (vector of size $N$) such that:
\begin{eqnarray}\label{eq:affine}
X = P.X + B.
\end{eqnarray}
where $P$ is a square matrix of size $N\times N$ and $B$ a vector of size $N$.
In particular, it has been shown how this iterative method can be further applied
to solve $X$ such that
$$Q.X = X \mbox{ and } R.X = B$$
where $Q$ and $R$ are square matrices of size $N\times N$ or to solve
$$A.X = B$$
where $A$ is a square matrix of size $N\times N$.

We recall that the D-iteration approach works when the spectral radius
of $P$ is strictly less than 1 and that it basically consists in computing
efficiently the solution $X$ of the equation \eqref{eq:affine} using the
power series $X = \sum_{n=0}^{\infty} P^n B$.

\section{Equation on $H_n$}\label{sec:history}
The fluid diffusion model is in the general case described
by the matrix $P$ associated with a weighted graph ($p_{ij}$ is
the weight of the edge from $j$ to $i$) and the initial condition
$F_0 = B$.

We recall the definition of the two vectors used in D-iteration:
the fluid vector $F_n$ defined by:
\begin{eqnarray}
F_n &=& (I_d - J_{i_n} + P J_{i_n}) F_{n-1}.\label{eq:defF}
\end{eqnarray}
where:
\begin{itemize}
\item $I_d$ is the identity matrix;
\item $I = \{i_1, i_2, ..., i_n,...\}$ with $i_n \in \{1,..,N\}$ is a deterministic or random sequence
  such that the number of occurrence of each value $k\in \{1,..,N\}$ in $I$ is infinity;
\item $J_k$ a matrix with all entries equal to zero except for
  the $k$-th diagonal term: $(J_k)_{kk} = 1$.
\end{itemize}

And the history vector $H_n$ defined by ($H_0$ initialized to a null vector):
\begin{eqnarray}\label{eq:defH}
H_n &=& \sum_{k=1}^n J_{i_k} F_{k-1}.
\end{eqnarray}

Then, we have (cf. \cite{dohy}):
\begin{eqnarray}\label{eq:H}
H_n + F_n &=& F_0 + P H_n.
\end{eqnarray}

It has been shown in \cite{dohy} that $H_n$ satisfies the equation:
\begin{eqnarray}\label{eq:hlin}
H_{n} &=& \left(I_d - J_{i_n}(I_d - P)\right)H_{n-1} + J_{i_n} F_0.
\end{eqnarray}

In fact, the above equation can be very easily understood remarking that
$I_d - J_{i_n}(I_d - P)$ is a matrix built from $P$ extracting the
$i_n$-th line of $P$ and completing the rest with identity line vectors
on $i \neq i_n$ (zero everywhere except the $i$-th column equal to one).

Note that for the entry $i\neq i_n$, $(H_n)_i = (H_{n-1})_i$.

\subsection{Preliminary operations}
\subsubsection{Initial condition}
It is easy to see from the equation \eqref{eq:hlin} that when we choose $i_1=1, i_2=2,.., i_N=N$,
we obtain $H_N = B$. So we can directly start the iteration with $H_0=B$ without
any cost.

\subsubsection{Diagonal link elimination}
Now we can optionally apply the diagonal link elimination based on the method
defined in \cite{d-algo}: when $p_{ii} \neq 0$ is to be suppressed, it implies two modifications:
\begin{itemize}
\item modification of the initial fluid: replace $B_i$ by $B_i/(1-p_{ii})$;
\item modification of all link weights pointing to node $i$ (incoming links to $i$, namely all $j$ such that
  $p_{ij}\neq 0$): this operation
  can be replaced by keeping locally at node $i$ the information that all incoming
  fluid need to be multiplied by $1/(1-p_{ii})$.
\end{itemize}

\section{Distributive computation}\label{sec:distributive}
In the following we set $L_i(P)$ the $i$-th line vector extracted from $P$:
$$
(L_i(P))_j = p_{ij}.
$$
We start by assuming that there is a partition of $N$ in $K$ disjoint sets
$\Omega_i$, $i=1,..,K$, such that $\cup_{k=1}^K \Omega_k = \{1,..,N\}$.

The choice of the partition can be seen as an independent optimization task that will not
be discussed here (intuitively, $\Omega_k$ should be such that most of links are between
nodes of the same set).

\subsection{Operations in $\Omega_k$}\label{sec:oper}
We assume here that all computations of $(H_n)_i, i\in \Omega_k$ is handled
by one independent process (or server or virtual machine), that we call $PID_k$.

$PID_k$ has as input $B$ and $H$. $H$ is initially set to $B$.
\subsubsection{Local updates}
$PID_k$ updates $H$ by applying the fluid diffusion model with $i_n\in\Omega_k$:
\begin{eqnarray}\label{eq:up}
(H)_{i_n} = L_{i_n}(P).H + (B)_{i_n}.
\end{eqnarray}
\subsubsection{Updates sharing}
Periodically, $PID_k$ sends to all other $PID_i$ ($i\neq k$) the updated $(H)_{j\in\Omega_k}$.
When, a $PID_k$ receives updates of $(H)_i$ for $i\in \Omega_{k'}$, it updates the
current $H$ and can apply the local updates \eqref{eq:up}.

\subsection{Evolution of $P$}
If for some reason, the matrix $P$ is updated to a new matrix $P'$
and if one is interested by the solution of \eqref{eq:affine} with $P'$,
the new $P'$ is sent to all $PID_k$ that are concerned by the modification.

Upon reception of this modification, each $PID_k$ does the following updates:
\begin{itemize}
\item store the last result $H$ for entries $i\in\Omega_k$ (can be used as the
  new initial vector $H_0'$);
\item replace $B$ by $B' = F + (P'-P)H$ for entries $i\in\Omega_k$.
\end{itemize} 

$(F)_i$ is computed by: $L_{i}(P).H + (B)_{i} - (H)_i$.

Since each $PID_k$ only requires the information $(B)_i$ for $i\in\Omega_k$,
we don't need to synchronize for the new $B'$, but just update $B'$ locally
and then we can re-apply the methods of Section \ref{sec:oper} with $P'$.

The above result is based on the result of Theorem 4 of \cite{dohy}.

\subsection{Another version based on two state vectors (V2)}
The drawback of the above method is to have to keep the complete $H$ vector
for each PID.
For a really very large matrix $P$ this may be an issue.
In such a case, we may use the two fluid diffusion state vectors
$H_n$ and $F_n$ (equations \eqref{eq:defH} and \eqref{eq:defF}).
Then each $PID_k$ needs to keep only locally the partial view:
$(B)_i$, $(H_n)_i$ and $(F_n)_i$ only for $i\in\Omega_k$.

In such a scheme, the exchanged information between PIDs is the quantity
$F_n$ that need to be sent/received: each $PID_k$ exploits the column vector
extracted from $P$, say $C_i(P)$ for the $i$-th column vector ($i\in\Omega_k$).
When the diffusion is applied on node $i_n\in\Omega_k$ with the fluid
$f = (F_{n-1})_{i_n}$, the quantity $f\times p_{j i_n}$ need to be sent
to a $PID_{k'}$ such that $j\in\Omega_{k'}$, so that $PID_{k'}$ can
add this quantity to $(F_{n'})_j$.

The fluid transmission ($f\times P_{j i_n}$ to all $j$) does not require
any synchronization. To avoid too much information exchange, the fluid
transmission can be delayed and regrouped (we can regroup
$(f_1+f_2+..+f_m)\times P_{j i_n}$ so that this quantity is not too small;
we can regroup on $i_n$ as well if going to the same destination $j$):
in fact, we don't need to know who sent the fluid.
The only constraint is that the fluid transmission is not lost:
this means that each $PID_k$ need to keep locally
the information of the fluid  $(f_1+f_2+..+f_m)\times P_{j i_n}$ until
its destination PID ($PID_{k'}$) acknowledges its reception (say as TCP).

In this scheme, the convergence is explicitly monitored by observing the
total fluid quantity (locally updated $F_n$ plus all fluids being transmitted).

\section{Optimization problem}\label{sec:op}
Given the partition set $\Omega_k$, the question is when to share
the local updates on $H$. Here is a first possible solution.

\subsection{Local remaining fluid}
We can define the local remaining fluid $r_k$ by: 
$$r_k = \sum_{i\in \Omega_k} |L_{i}(P).H + (B)_{i} - (H)_i|.$$
Assuming a non-negative matrix $P$ and applying ideas of \cite{dohy},
we could decide to share the results of the local computations
to other PIDs when 
$$
r_k < T_k
$$ 
where $T_k$ is the local threshold for $\Omega_k$.
When such a condition is satisfied, we could then apply
an update of $T_k$.
For instance by a multiplicative division by factor $\alpha>1$:
$$
T_k := T_k/\alpha.
$$

In the version (V2), $r_k$ is explicitly given by the norm $L_1$ of $F_n$:
$r_k = \sum_{i\in\Omega_k} |(F_n)_i|$.

\subsection{Diffusion sequence $I$}
Here we need to choose the sequence order $i_n\in\Omega_k$ for each $k$.
By default, we can apply a cyclic order. We could apply also some greedy
approach as in \cite{dohy, d-algo}. 
Finding the optimal sequence or a practical sub-optimal
sequence for each $k$ is an open problem.

\subsection{Sharing locally updated results}
The transmission of $H$ to other PIDs is triggered when
\begin{itemize}
\item $r_k < T_k$, or
\item an update of $H$ is received from another PID.
\end{itemize}

In the version (V2), $F$ may be sent only when:
\begin{itemize}
\item $r_k < T_k$.
\end{itemize}

When the PIDs advance at very different speeds (monitoring $T_k$),
we can think of splitting the set $\Omega_k$ associated to the slowest $PID_k$
or possibly regrouping $\Omega_k$ associated to the fastest $PID_k$ etc.

\subsection{Distance to the limit}
The limit is reached when $\sum_k r_k = 0$.
In case of PageRank style equations, it has been
shown in \cite{dohy} that $(\sum_k r_k)/(1-d)$ defines an exact
distance to the limit or an upper bound in the presence of dangling nodes.

In the general case, the spectral radius of $P$ plays a role
(but is not necessarily known). For instance, if for all $i$,
$\sum_j |p_{ji}| < 1$, then taking $\epsilon = \min_i (1 - \sum_j |p_{ji}|)$,
$(\sum_k r_k)/\epsilon$ defines an upper bound of the distance to the limit.

\section{Examples}\label{sec:exm}

\subsection{Example with 2 PIDs}
Let's take a simple example to illustrate the above method.
We set:
$$
A(1) =
\begin{pmatrix}
5&	3&	0&	0\\
3&	7&	0&	0\\
0&	0&	8&	4\\
0&	0&	2&	3
\end{pmatrix}
$$
And we look for $X$ such that $A.X = B = (1,1,1,1)^t$.

In this case, we defined $A(1)$ so that they is no correlation between
$\Omega_1 = \{1,2\}$ and $\Omega_2 = \{3,4\}$. As expected, then the
gain factor is about 2 (assuming no information transmission cost) with
2 PIDs as shown
in Figure \ref{fig:compa1}: in Figure \ref{fig:compa1}, we compared the Jacobi
and Gauss-Seidel iterations and the D-iteration on $P$ obtained from $A$
by dividing each line by the diagonal term (cf. \cite{d-algo}):
$$
P =
\begin{pmatrix}
0&	-3/5&	0&	0\\
-3/7&	0&	0&	0\\
0&	0&	0&	-4/8\\
0&	0&	-2/3&	0
\end{pmatrix}
$$
For the D-iteration,
we applied the cyclical sequence $\{1, 2, 3, 4\}$ (using the equation \eqref{eq:hlin} on $H_n$). 
For 2 PIDs case, we applied
jointly the cyclical sequence $\{1, 2\}$ and $\{3, 4\}$ exactly twice before sharing
the local computation results. 

\begin{figure}[htbp]
\centering
\includegraphics[angle=-90,width=\linewidth]{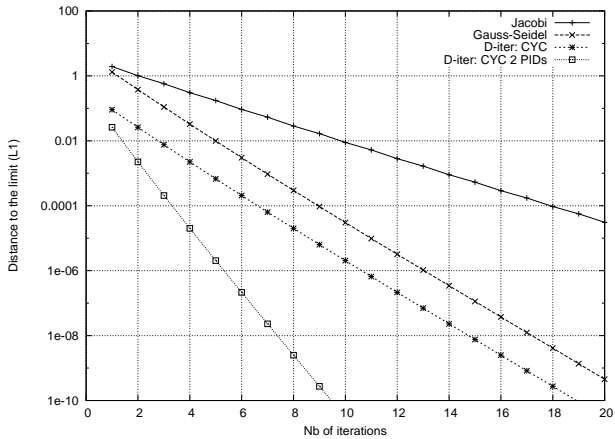}
\caption{Example: 2 PIDs for $A(1)$.}
\label{fig:compa1}
\end{figure}

Now, we set:
$$
A(2) =
\begin{pmatrix}
5&	3&	1&	1\\
3&	7&	1&	0\\
1&	1&	8&	4\\
1&	1&	2&	3
\end{pmatrix}
$$

In this case, we added values in $A(2)$ so that they is correlation between
$\Omega_1$ and $\Omega_2$. 
Then there is still a visible gain factor as shown
in Figure \ref{fig:compa2}.

\begin{figure}[htbp]
\centering
\includegraphics[angle=-90,width=\linewidth]{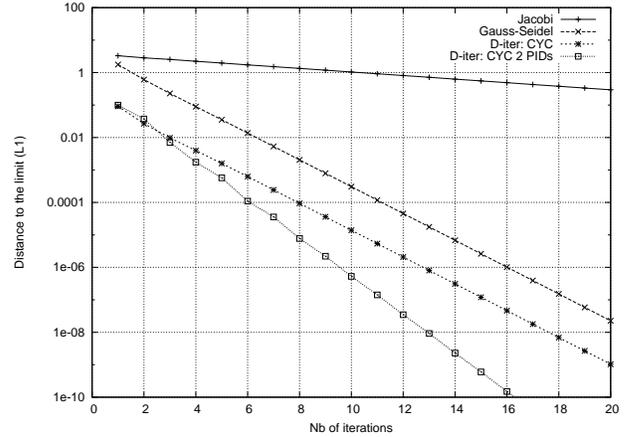}
\caption{Example: 2 PIDs with correlation for $A(2)$.}
\label{fig:compa2}
\end{figure}

Finally, we set:
$$
A(3) =
\begin{pmatrix}
5&	3&	1&	1\\
3&	7&	1&	1\\
1&	1&	8&	4\\
1&	1&	2&	3
\end{pmatrix}
$$

In this case, we added 1 on $(A(3))_{2,4}$. 
Then there is no longer any significant gain as shown
in Figure \ref{fig:compa3}.

\begin{figure}[htbp]
\centering
\includegraphics[angle=-90,width=\linewidth]{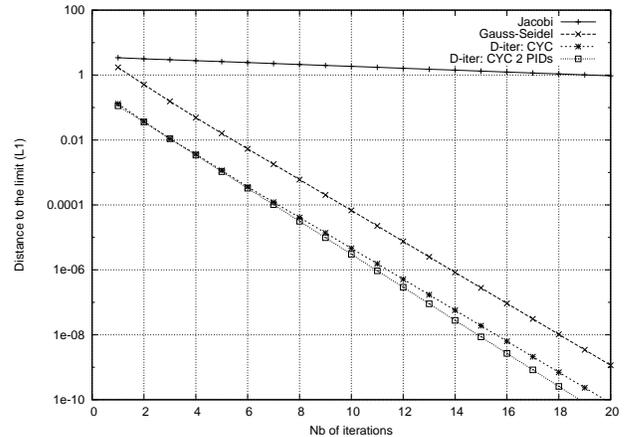}
\caption{Example: 2 PIDs with correlation for $A(3)$.}
\label{fig:compa3}
\end{figure}

\subsection{Example of $A$ updates with 2 PIDs}

We set:
$$
A =
\begin{pmatrix}
5&	3&	0&	0\\
3&	7&	0&	0\\
0&	0&	8&	4\\
0&	0&	2&	3
\end{pmatrix}
$$
and
$$
A' =
\begin{pmatrix}
5&	3&	0&	0\\
3&	7&	0&	1\\
0&	0&	8&	4\\
0&	0&	2&	3
\end{pmatrix}
$$

Then $P$ and $P'$ are defined by:
$$
P =
\begin{pmatrix}
0&	-3/5&	0&	0\\
-3/7&	0&	0&	0\\
0&	0&	0&	-4/8\\
0&	0&	-2/3&	0
\end{pmatrix}
$$
and
$$
P' =
\begin{pmatrix}
0&	-3/5&	0&	0\\
-3/7&	0&	0&	-1/7\\
0&	0&	0&	-4/8\\
0&	0&	-2/3&	0
\end{pmatrix}
$$

$P$ has been applied up to iteration 5, then we switched to $P'$ from iteration 6.
Figure \ref{fig:compa-update} shows the results: 

\begin{figure}[htbp]
\centering
\includegraphics[angle=-90,width=\linewidth]{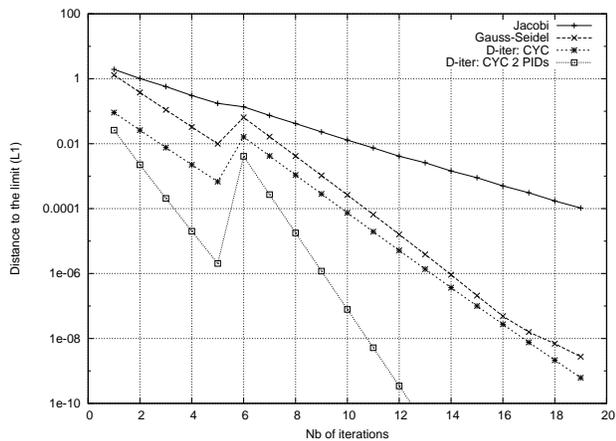}
\caption{Example: 2 PIDs with evolution of $P$ to $P'$.}
\label{fig:compa-update}
\end{figure}

The above examples are only for easy illustration. The gain of the distributed approach
should be much clearer for the computation of $X$ for large matrix $P$.
This will be addressed in a future paper in the context of the PageRank equations,
on the web graph (on which the gain of such an approach without distributed computations
is shown in \cite{dohy}) or on the general graph (such as the PageRank extensions on
the paper-author graph for the research publications \cite{pira}).

\section{Conclusion}\label{sec:conclusion}
In this paper, we presented two asynchronous computation schemes associated to the
D-iteration approach. We believe that its potential is very promising and further 
investigation (and implementation)
for a really large $P$, such as for the PageRank matrix associated to the web graph, will
be addressed in a future paper.

\section*{Acknowledgments}
The author is very grateful to G\'erard Burnside for his
valuable comments and suggestions.
\end{psfrags}
\bibliographystyle{abbrv}
\bibliography{sigproc}

\end{document}